\documentclass[letterpaper,12pt]{article} 

\usepackage{hyperref}
\usepackage{amssymb}
\usepackage{amsmath}
\usepackage{graphicx}
\usepackage{xcolor}

\title{A polynomial regression model for excess mortality in Mexico 2020-2022 due to the COVID-19 pandemic}
\author{Andreu Comas-Garc\'ia\footnote{\textbf{andreu.comas@uaslp.mx} Deparment of Microbiology, School of Medicine, Universidad Aut\'onoma de San Luis Potos\'i and Universidad Cuauht\'emoc Plantel San Luis Potos\'i (M\'exico)}, Arturo Erdely\footnote{\textbf{aerdely@acatlan.unam.mx} Facultad de Estudios Superiores Acatl\'an, Universidad Nacional Aut\'onoma de M\'exico}}

\begin{document}
	
\maketitle

\begin{abstract}
    Based on the comprehensive national death registry of Mexico spanning from 1998 to 2022 a point and interval estimation method for the excess mortality in Mexico during the years 2020-2022 is proposed based on illness-induced deaths only, using a \textit{polynomial regression} model. The results obtained estimate that the excess mortality is around 788,000 people (39.3\%) equivalently to a rate of 626 per 100,000 inhabitants. The male/female ratio is estimated to be 1.7 times. As a reference for comparison, for the whole period 2020-2020 Mexico's INEGI estimated an excess of mortality between 673,000 with a quasi-Poisson model and 808,000 using endemic channels estimation.
\end{abstract}



\section{Introduction}	

Assessing excess mortality is crucial for gaining a comprehensive understanding of the pandemic's impact \cite{Beaney2020}. Excess deaths encompass not only those directly attributed to SARS-CoV-2 infection but also the indirect consequences, including the strain on the healthcare system and the effects of mitigation measures. To estimate excess deaths, one must establish a pre-pandemic reference period, employ a model to project expected deaths during the pandemic, and then compare these projections with the observed deaths \cite{Levitt2022}. \medskip

Mexico, alongside nations like India, United States of America, Russia, Brazil, Indonesia, and Pakistan globally, and Brazil, Chile, Ecuador, Guatemala, and Peru within Latin America, has experienced one of the highest levels of excess mortality during the pandemic \cite{Karlinsky2021}. The initial forecasts predicted a range of excess death rates, spanning from 45\% to 52\% in 2020. Mexico was expected to witness an additional 798,000 deaths by December 2021, placing it among the top seven countries worldwide for excess mortality. However, \cite{Palacio2022} estimated that excess mortality was lower than predicted, with a 38.2\% increase in deaths due to all causes during 2020 and 2021. \medskip

In contrast to some other nations where the elderly bore the majority of COVID-19 mortality, Mexico displayed a distinctive pattern with a higher death rate observed among individuals aged 45 to 64. Factors such as a high prevalence of comorbidities at younger ages, including diabetes, hypertension, and obesity --known risk factors for COVID-19 mortality-- may have contributed to this distinct age-related pattern in Mexico \cite{Garcia2021}. \medskip

In this study, our primary goal was to investigate and quantify the temporal distribution of excess mortality attributed to the COVID-19 pandemic in Mexico across various sex-age groups from 2020 to 2022. By examining these temporal changes, we aimed to understand the impact of the pandemic on each group. Under the conducted analysis it was found that a \textit{polynomial regression} model was a realistic fit for the pre-pandemic yearly per week distribution of illness-related deaths. It is proposed to estimate the excess mortality after subtracting all deaths that were not related to a disease, for example, deaths due to violence or accidents, following \cite{WHO2023}:
\begin{quote}
    \textsl{When calculating excess deaths, it is crucial to differentiate between deaths directly or indirectly attributable to COVID-19 and deaths resulting from other non-disease causes, such as those related to conflicts, violence, accidents, and natural disasters. Hence, it is a common practice to exclude deaths from other shocks when determining the total number of excess deaths. This approach aids in obtaining a more accurate estimate of the impact of COVID-19 on mortality.}
\end{quote}
The proposal also considers mortality trends in Mexico from previous pre-pandemic years \cite{Islam2022}. The results thus obtained are between two official INEGI estimations \cite{INEGI2023a} that use other approaches (endemic channels and a quasi-Poisson model).


\section{Data}

Our analysis relies on the extensive national death registry database of Mexico, covering the period from 1998 to 2022 \cite{INEGI2023} and \cite{DGIS2023}. Although the majority of deaths are typically recorded within the same calendar year in which they occur, there are exceptions to this temporal alignment. Specifically, 2.6\% of deaths registered in 2022 were reported as having occurred in previous years. Additionally, some deaths that took place in 2022 will be included in the 2023 database, particularly those that transpired close to the end of the year.\medskip

To ensure the accurate analysis of mortality based on the actual date of death, it is crucial to address discrepancies between the year of death and the year of registration during the process of data extraction and filtration. This involves meticulous attention to detail and the implementation of appropriate adjustments to account for cases where the registration year does not align with the actual year of death.\medskip

To enhance consistency in year-to-year comparisons and to align more effectively with our subsequent modeling approach, we have chosen a standardized definition of weeks rather than relying on the epidemiological weeks recommended by the Pan American Health Organization (PAHO) \cite{PAHO2016}. This decision stems from the variability in the number of epidemiological weeks per year, which can be either 52 or 53, and their occasional spanning of two different calendar years. \medskip

In our proposed methodology, week 1 is designated as the initial seven days of the year, week 2 encompasses days 8 through 14, and this pattern continues up to week 52, covering days 358 to 364. To account for the remaining days of the year, we include an additional always incomplete ``week'' 53, which consists of only day 365 for common years and both days 365 and 366 for leap years. This approach has been adopted in lieu of the alternative method, which involves dividing the number of deaths in week 53 by 2, adding half to week 52, and the remaining half to week 1 of the subsequent year. \medskip

While a conventional epidemiological approach for estimating excess mortality during an epidemic involves calculating the disparity between observed all-cause death counts and the anticipated counts in the absence of an epidemic (see for example \cite{PAHO2020}), we deviate from this method by excluding deaths unrelated to illnesses. This departure aligns with the recommendation from \cite{WHO2023}, which suggests the exclusion of deaths resulting from other shocks. This refined approach aims to enhance the accuracy of estimating the impact of COVID-19 on mortality by isolating and focusing specifically on deaths directly or indirectly attributable to the virus, thereby providing a more precise assessment of its effect on overall mortality rates.\medskip 

\begin{figure}[h]
	\centering
	\includegraphics[width=8cm]{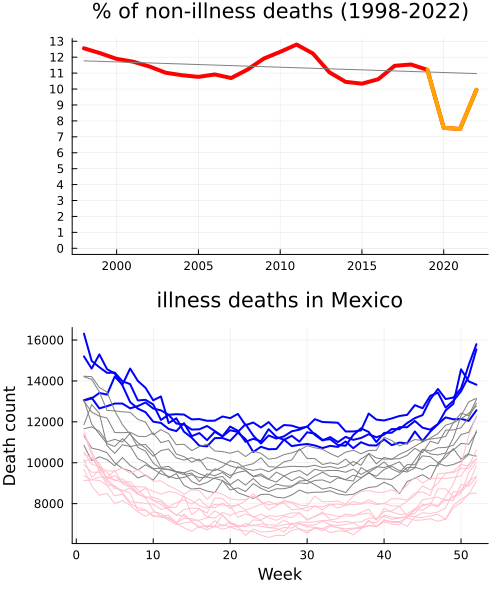}
	\caption{a) Top: Percentage of non-illness deaths per year in Mexico during the period 1998-2022. b) Bottom: Number of illness-related deaths per week for each year in Mexico, with individual polygonal curves representing the data for 52-week spans, during the periods 1998-2007 (pink), 2008-2015 (gray), and 2016-2019 (blue).}
	\label{fig:prcnoenf_enf9819}
\end{figure}

Figure \ref{fig:prcnoenf_enf9819}a distinctly demonstrates that in the year preceding the COVID-19 outbreak in Mexico, slightly over 11\% of total deaths were attributed to non-illness causes, such as accidents or violence. This proportion exhibited a noticeable decline to 8.5\% during the first two years of the epidemic. This reduction is likely a consequence of widespread lockdowns and substantial restrictions on social and economic activities, occurring concurrently with a notable increase in deaths related to illnesses. \medskip 

Nevertheless, it is crucial to highlight that this percentage is swiftly converging towards pre-pandemic levels in 2022. The resurgence implies a potential rebound in non-illness-related fatalities, hinting at the resumption of normal social and economic activities. This observation underscores the dynamic interplay between public health measures, societal restrictions, and the evolving composition of mortality causes during and after the peak of the COVID-19 pandemic. It points to the complex and multifaceted influences shaping mortality patterns as societies navigate the aftermath of significant public health events. \medskip 

As a result, we advocate for a calculation method that contrasts observed illness-induced deaths with expected illness-induced deaths exclusively. Our approach does not solely focus on deaths directly attributed to COVID-19 for several reasons: a) The novelty of the disease meant it could be misidentified as other illnesses; b) COVID-19 could exacerbate underlying health issues, leading to fatality; c) The epidemic's surge in demand for healthcare services might have meant that other illnesses requiring specialized care were potentially neglected, indirectly contributing to an increased mortality rate; and d) The national hospital re-conversion strategy could have influenced the management of non-COVID-19 medical illnesses. These factors underscore the necessity of including all illness-related deaths during the epidemic period in our analysis, providing a more comprehensive understanding of the broader impact on mortality beyond direct COVID-19-related fatalities. \medskip

To calculate mortality rates per $100,000$ inhabitants, we utilized population projections provided by CONAPO \cite{CONAPO2023}. However, it's important to note that these projections were adjusted based on the results of Mexico's national population census conducted in 2020 by \cite{INEGI2020}. This adjustment ensures a more accurate representation of the population dynamics and enhances the precision of our mortality rate calculations in reflecting the actual demographic landscape as revealed by the latest census data.


\section{Methods} 

Figure \ref{fig:prcnoenf_enf9819}b illustrates the number of illness-related deaths per week for each year, showcasing individual polygonal curves representing data over 52-week spans. For clarity in visualization, data from the always incomplete 53rd week (as explained in the previous section) has been omitted in the graph. However, it's important to note that this excluded data is factored into the total excess mortality calculations, which we will elaborate on subsequently. Two significant patterns emerge from the graph: 1) an annually consistent U-shaped trend, and 2) a noticeable gradual increase in deaths over time. The latter can be partially attributed to Mexico's ongoing population growth, recorded at a rate of 1.03\% in 2019 \cite{CONAPO2023}. \medskip

The distribution of weekly deaths reveals a U-shaped pattern, indicating that illness-related fatalities are not evenly spread throughout the year. Instead, there is an elevated occurrence of deaths in the winter months, peaking at the beginning and end of each year. Conversely, the lowest numbers are consistently recorded between weeks 20 and 40, with only minor fluctuations throughout this period. \medskip 

To accurately model this seasonal pattern, we have chosen to employ an even-degree polynomial. While a quadratic polynomial with a positive coefficient of the quadratic term can create a U-shaped curve, it lacks the flexibility to capture subtle fluctuations around the annual minimum. On the other hand, a fourth-degree polynomial provides the capacity to reflect these minor oscillations. Polynomials of higher degrees often result in excessive, unrealistic oscillations and introduce additional parameters that can amplify noise, potentially complicating the model. \medskip

We have identified that a fourth-degree polynomial strikes a balance between a realistic representation of the data and avoiding the pitfall of overfitting. The proposed model for observed deaths is thus defined as:

\begin{equation}\label{eq:model}
	d(w) \,=\, \alpha + \beta_{1}w + \beta_{2}w^2 + \beta_{3}w^3 + \beta_{4}w^4 + \varepsilon 
\end{equation}
where week $w$ takes values $\{1,2,\ldots,52\}$ and $\varepsilon$ is a random noise variable with the only requirements that its mean value equals to zero, and finite variance. We should notice that 52 weeks of 7 days each account for 364 days of the year. Parameters $\alpha,$ $\beta_{1},$ $\beta_{2},$ $\beta_{3},$ and $\beta_{4}$ are estimated by the ordinary least squares (OLS) techinque (see for example \cite{Weisberg2005}). With the estimated parameters we get an OLS point estimation of $d(w)$ which is denoted by:

\begin{equation}\label{eq:fitted_model}
	\widehat{d}(w) \,=\, \widehat{\alpha} + \widehat{\beta_{1}}w + \widehat{\beta{2}} + \widehat{\beta_{3}}w^3 + \widehat{\beta_{4}}w^4
\end{equation}
The incomplete 53rd week may be estimated as follows:

\begin{equation}\label{eq:week53}
	\widehat{d(53)} \,=\, \widehat{d}(53) \times \frac{\ell}{7} \times b
\end{equation}
where $\ell=1$ for non-leap years (day 365) and $\ell=2$ for leap years (days 365 and 366), and $b$ is a bias adjustment factor calculated as the average value of:

\begin{equation}\label{eq:bias}
	\frac{d(53)}{\widehat{d}(53)\times\frac{\ell}{7}}
\end{equation}
over the period 1998-2019 in Mexico. A linear trend is calculated for the estimated parameters, adding a parameter $\sigma$ for the standard deviation of the residuals $\hat{\varepsilon}_{1},\ldots,\hat{\varepsilon}_{52}$ given by $\hat{\varepsilon}_{w}=\widehat{d}(w)-d(w)$ for $w$ in $\{1,2,,\ldots,52\}.$ The values for 2020, 2021, and 2022 are imputed from the linear trend of the parameters during the 1998-2019 period. \medskip 

Model (\ref{eq:fitted_model}) is fitted for each of the years 1998 to 2019, a total of 22 cases. For each case the goodness-of-fit was checked by calculating the adjusted R-squared and the Anderson-Darling's \cite{Anderson1954} normality test p-value for the residuals, and in all cases high R-squared values account for a good explanatory model (all above $0.8$), and also p-values far away from the usual $0.05$ threshold (all above $0.45$) indicate that a normality assumption for the random error $\varepsilon$ is far from being rejected. \medskip

It is usually difficult to find a useful interpretation of the coefficients in a polynomial regression model, except in this particular case for $\alpha$ which acts as a shift (or baseline) parameter for each year, and which exhibits a not surprising increasing trend since along with the (still) positive growth rate of Mexico's population, with more exposed people it is expected to have more illness-related deaths in absolute numbers. \medskip

Moreover, we may fit a linear trend for parameter $\alpha$ also by OLS:

\begin{equation}\label{eq:alpha}
	\alpha(t) \,\approx\, \widehat{\alpha}(t) = mt + c\,,\quad t \in \{1998,\ldots,2019\}
\end{equation}
and we may calculate the percentage increase rate of such shift parameter $\alpha$ for each year from 1998 until 2019:

\begin{equation}\label{eq:Deltaalpha}
	\widehat{\Delta\alpha}(t) \,=\, \frac{m}{mt + c}\,\times\,100\%\,,\quad t\in \{1998,\ldots,2019\}
\end{equation}
and compare the latest just before the beginning of the COVID-19 pandemic to Mexico's population annual growth rate in 2019 which was 1.03\% in contrast with $\widehat{\Delta\alpha}(2019)=1.82\%$ showing that illness-related deaths in Mexico were growing faster than its population, and therefore an ongoing health deterioration of health in Mexico was already taking place right before de pandemic. \medskip

Finally, we may estimate the total excess mortality ($\Psi$) for the epidemic years 2020, 2021, and 2022, as the difference of total illness deaths observed in each period minus the expected illness-related deaths according to the fitted model (\ref{eq:fitted_model}) with a forecast of the parameters using their linear trends: 

\begin{equation}\label{eq:totexcess}
	\Psi \,=\, \sum_{w\,=\,1}^{52} \left[\,d(w)-\widehat{d}(w)\,\right] + d(53) - \widehat{d(53)}
\end{equation}
And also the total percentage excess mortality as:

\begin{equation}\label{eq:totexcessprc}
	\Delta\Psi \,=\, \frac{\Psi}{\sum_{w=1}^{52}\widehat{d}(w) + \widehat{d(53)}}\,\times\,100\%
\end{equation} \medskip

The calculation of mortality rates per $100,000$ inhabitants is then calculated as the absolute total excess mortality (\ref{eq:totexcess}) divided by the estimated population size, using the projections by CONAPO \cite{CONAPO2023} but updated with the results of the population census by INEGI \cite{INEGI2020}.


\section{Results}

With the methods proposed in the previous section and the available data it was possible to fit model (\ref{eq:fitted_model}) for different aggregation levels of interest for the present work, in terms of period, sex and age groups, and estimate the illness excess mortality for each case: 
\begin{enumerate}
	\item Periods: 2020, 2021, 2022, and 2020-2022
	\item Age groups: 0-5, 6-10, 11-19, 20-29,\ldots, 60-69, 70+
	\item Sex: male, female, both
\end{enumerate}

As a result, we conducted 120 different estimations using model (\ref{eq:fitted_model}), considering 4 $\times$ periods, 10 age groups, and 3 sex categories. For each case, formulas (\ref{eq:totexcess}) and (\ref{eq:totexcessprc}) were applied to calculate excess mortality. The results are summarized in Tables \ref{tab:2020}, \ref{tab:2021}, \ref{tab:2022}, and  \ref{tab:202122} providing point and interval estimations for the years 2020, 2021, 2022, and the combined period of 2020-2022, respectively. Further, Tables \ref{tab:202122hom} and  \ref{tab:202122muj} present a breakdown by sex (male, female), summarizing estimations for the entire period of 2020-2022. \medskip

\begin{table}[h]
	\centering
	\includegraphics[width=10cm]{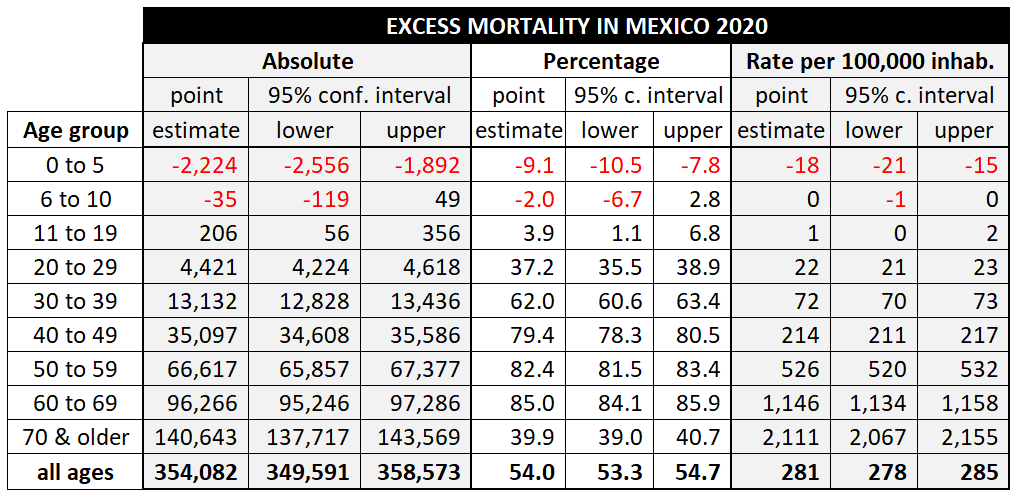}
	\caption{Estimated illness excess mortality in Mexico for the year 2020 in absolute numbers, percentage and rate per 100,000 inhabitants.}
	\label{tab:2020}
\end{table}
\begin{table}[h]
	\centering
	\includegraphics[width=10cm]{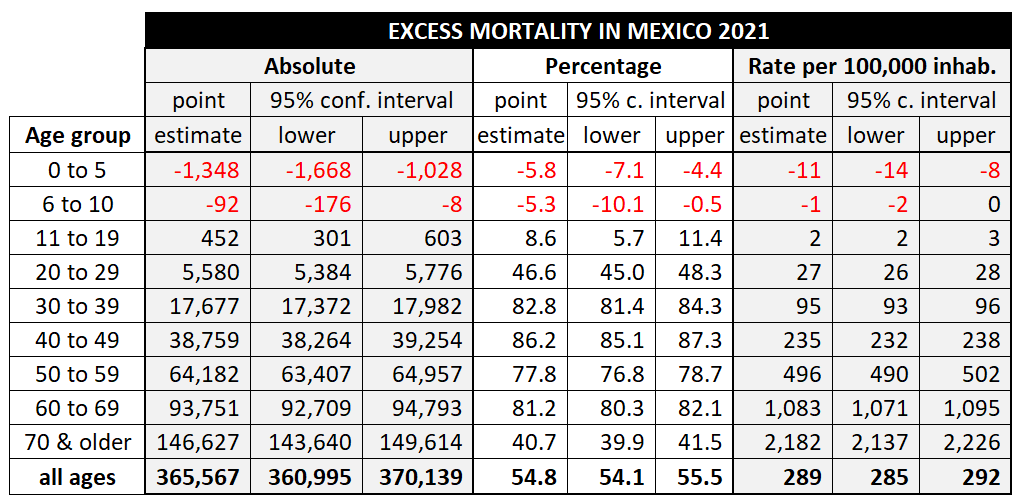}
	\caption{Estimated illness excess mortality in Mexico for the year 2021 in absolute numbers, percentage and rate per 100,000 inhabitants.}
	\label{tab:2021}
\end{table}
\begin{table}[h]
	\centering
	\includegraphics[width=10cm]{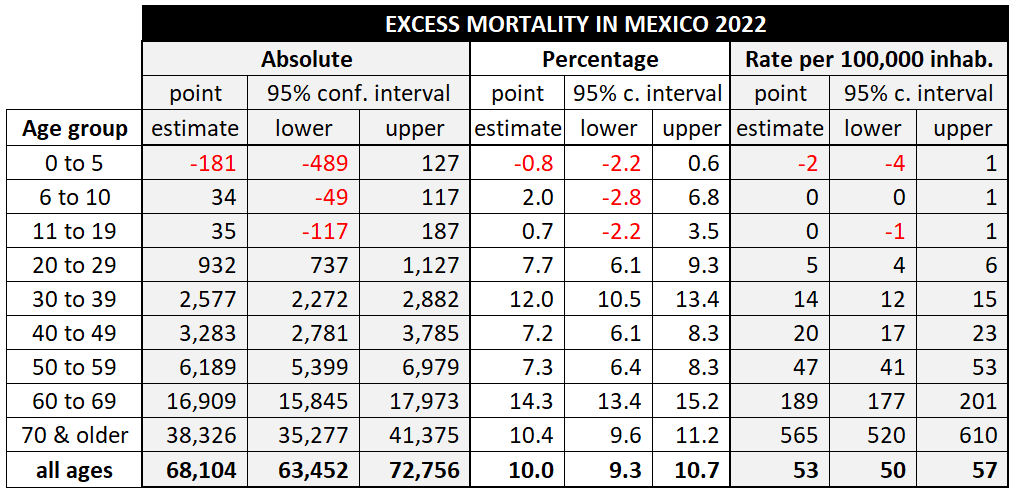}
	\caption{Estimated illness excess mortality in Mexico for the year 2022 in absolute numbers, percentage and rate per 100,000 inhabitants.}
	\label{tab:2022}
\end{table}
\begin{table}[h]
 	\centering
 	\includegraphics[width=10cm]{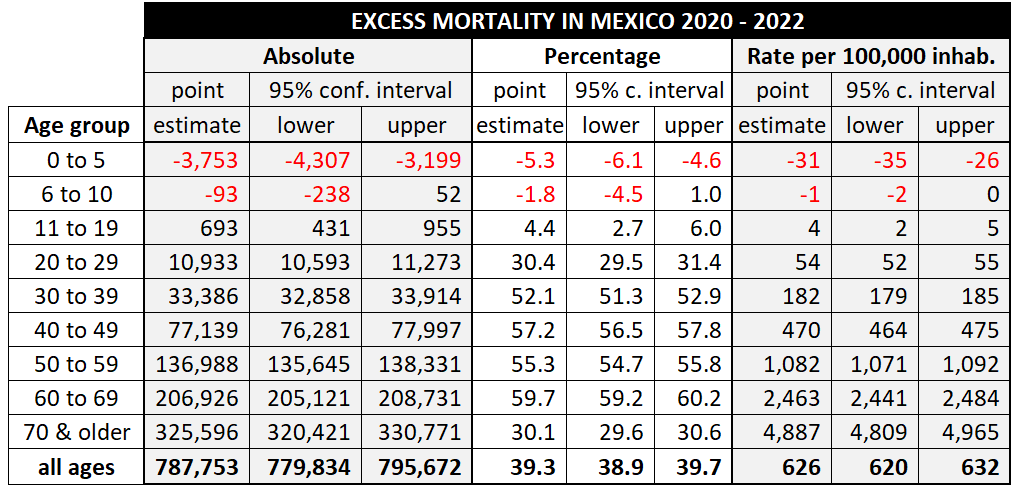}
 	\caption{Estimated illness excess mortality in Mexico for the three-year period 2020-2022 in absolute numbers, percentage and rate per 100,000 inhabitants.}
 	\label{tab:202122}
\end{table}
\begin{table}[h]
 	\centering
 	\includegraphics[width=10cm]{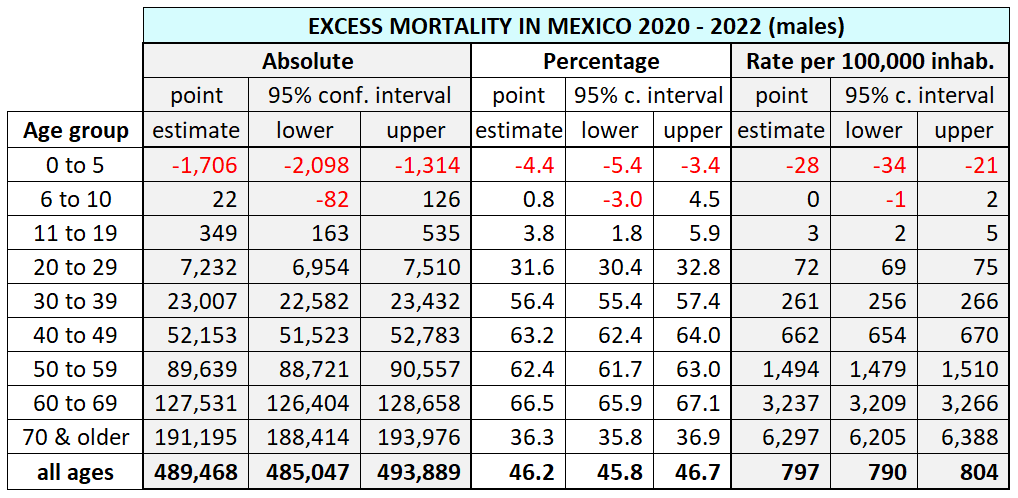}
 	\caption{Estimated illness excess mortality in Mexican males for the three-year period 2020-2022 in absolute numbers, percentage and rate per 100,000 inhabitants.}
 	\label{tab:202122hom}
\end{table}
\begin{table}[h]
 	\centering
 	\includegraphics[width=10cm]{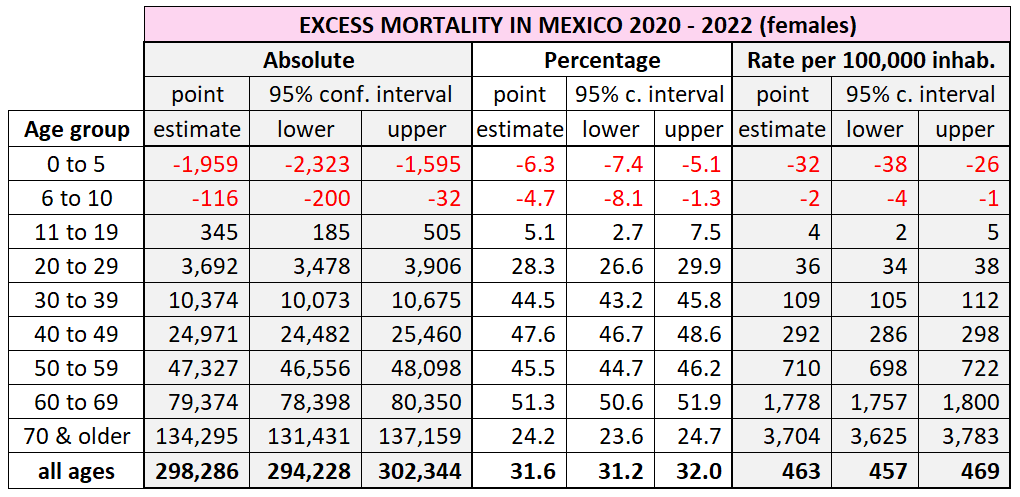}
 	\caption{Estimated illness excess mortality in Mexican females for the three-year period 2020-2022 in absolute numbers, percentage and rate per 100,000 inhabitants.}
 	\label{tab:202122muj}
\end{table}

\begin{figure}[h]
	\centering
	\includegraphics[width=10cm]{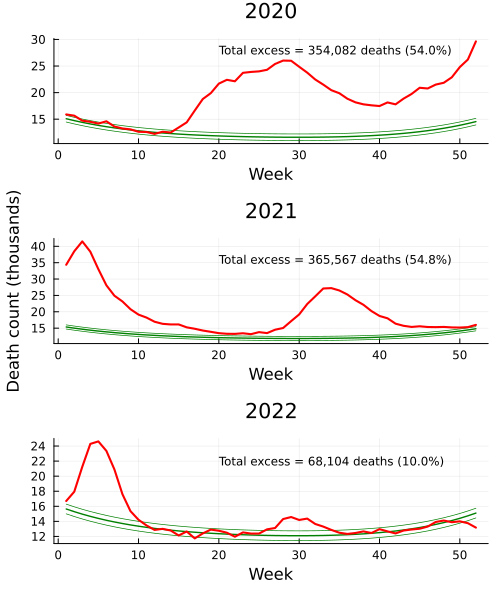}
	\caption{Observed (red) versus expected (green) illness-related deaths per week (with $95\%$ confidence bands) in Mexico during the years 2020, 2021, and 2022.}
	\label{fig:Exceso202122}
\end{figure}

In Figure \ref{fig:Exceso202122} , we present a comparison between observed and expected illness-related deaths per week for the years 2020, 2021, and 2022. Towards the end of 2022, the death counts appear significantly below the expected values. This discrepancy is primarily due to the unavailability of data from 2023 death registries, which will include fatalities that occurred at the end of 2022 but were registered in 2023. \medskip

\begin{figure}[h]
	\centering
	\includegraphics[width=10cm]{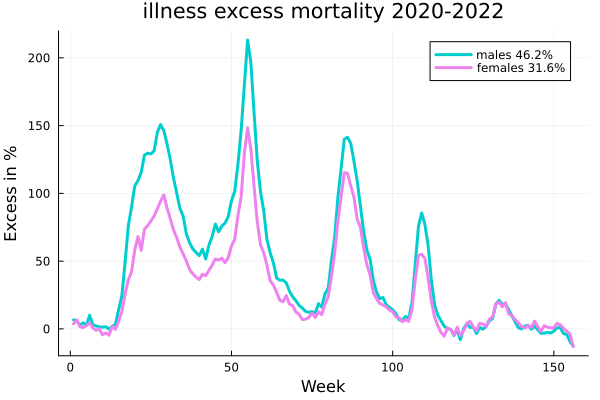}
	\caption{Males versus females percentage illness excess mortality in Mexico during the period 2020-2022.}
	\label{fig:ExcesoSexo}
\end{figure}

Moving on to \ref{fig:ExcesoSexo}, we depict a comparison of the percentage of illness-related excess mortality per week between males and females from 2020 to 2022. During the initial wave, there was a noticeable disparity in the proportion of excess deaths, with males experiencing a significantly higher rate compared to females. However, as subsequent waves unfolded, this gender difference gradually decreased over time. \medskip

For the initial year of the epidemic in Mexico (2020), the estimated illness-related excess mortality is $354,082$ deaths, with a $95\%$ confidence interval ranging from $349,591$ to $358,573$. This represents 54\% of the total expected mortality for the year or is equivalent to 281 excess deaths per 100,000 inhabitants, as shown in Table \ref{tab:2020}. Within the age group of $0$ to $5$ years, a negative excess mortality of $-9.1\%$ is estimated. However, for the age group of 6 to 10 years, the extent of reduction is unclear, given that the $95\%$ confidence interval spans from $-6.7\%$ to $+2.8\%$, encompassing both positive and negative values. For individuals aged 11 and older, an excess in mortality was recorded. The ratio of illness-related excess mortality between males and females was $1.9$ times, indicating that males experienced a significantly higher rate of excess mortality compared to females during the examined period. \medskip

In the second year of the epidemic in Mexico (2021), the overall illness-related excess mortality rate remained nearly the same, reaching $289$ per $100,000$ inhabitants compared to $281$ in 2020 (please refer to Tables \ref{tab:2020} and \ref{tab:2021} for comparison). Within the age group of $0$ to $5$ years, again a negative excess mortality but of $5.8\%,$ and for the age group of $6$ to $10$ years also a similar negative excess mortality of $5.3\%,$ in both cases the $95\%$ C.I completely in negative values, so that the negative excess mortality for individuals under $11$ years old is statistically significant. For the remaining age groups, there was a reduction in the ratio of males to females' excess deaths, ranging between $1.5$ to $2.3$ times higher in males. This indicates a decrease compared to the previous year, where the ratio ranged from $1.9$ to $2.5$. This shift in the gender difference suggests a potential change in the impact of the epidemic on different demographic groups. \medskip

In the third year of the epidemic in Mexico (2022), there is a significant decrease in illness-related excess mortality compared to the previous two years, with the rate dropping to $53$ excess deaths per $100,000$ inhabitants (refer to Table \ref{tab:2022} for details). Although the overall excess mortality rate male/female ratio was $1.5$ times, for individuals between 20 and 59 years old, this ratio was significantly higher than in the previous two years, ranging between $2.5$ and $3.5$ times. For the entire three-year period (2020-2022) of the COVID-19 epidemic in Mexico, the estimated total is nearly $788,000$ excess deaths, accounting for $39.3\%$ of the total mortality. This is equivalent to $626$ excess deaths per $100,000$ inhabitants (as detailed in Table \ref{tab:202122}). The cumulative figures provide a comprehensive overview of the impact of the epidemic on mortality over the examined period. \medskip

\begin{figure}[h]
	\centering
	\includegraphics[width=10cm]{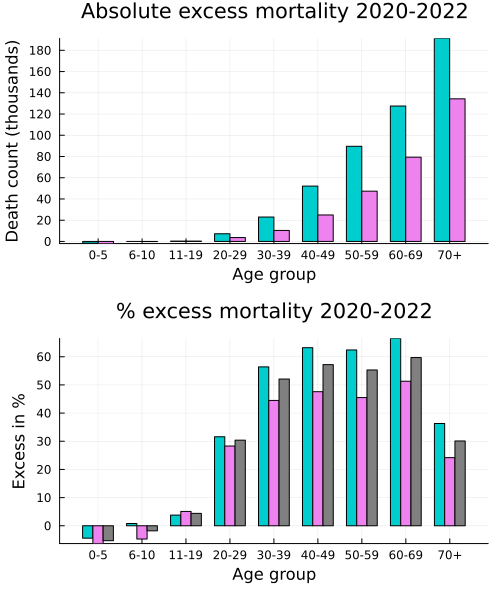}
	\caption{a) Top: Total illness excess mortality in Mexico during the period 2020-2022 by age groups and sex (males in blue, females in violet). b) Bottom: Percentage illness excess mortality in Mexico during the period 2020-2022 by age groups and sex (males in blue, females in violet, gray for both).}
	\label{fig:excesos202122}
\end{figure}

In absolute numbers, there is a discernible increase in illness-related excess mortality with advancing age in the population, as depicted in Figure \ref{fig:excesos202122}a. Regardless of age group, male mortality consistently surpassed that of females. Particularly noteworthy is the finding that excess mortality attributable to illness in men aged 60 to 69 was comparable to that observed in women over 70. This consistent pattern persisted across all age groups, emphasizing the similarity in female and male mortality rates but with an age difference of approximately a decade. This observation underscores the importance of age and sex as significant factors in understanding the impact of illness-related excess mortality. \medskip

When comparing age groups, it's evident that the elderly (70 years and older) endured the highest absolute number of fatalities. However, in terms of percentage excess mortality, this age group did not experience the most severe impact. Referring to Table \ref{tab:2020}, in 2020, the highest percentages were observed in the age range of 82.4\% to 85.0\% for individuals between 50 and 69 years old. In 2021 (as per Table \ref{tab:2021}), the worst cases shifted to younger people within the range of 82.8\% to 86.2\% for ages between 30 and 49. This shift can be considered an indirect consequence of the national COVID-19 vaccination strategy, which commenced with older individuals. The data underscores the evolving dynamics of the pandemic's impact on different age cohorts, influenced by vaccination roll-out strategies and other mitigating factors. \medskip 

From Table \ref{tab:2022} , it is evident that the percentage of illness-related excess mortality dropped significantly below 15\% for all age groups, indicating that the vast majority of the population had access to COVID-19 vaccination. Analyzing the cumulative data from 2020 to 2022 (refer to Table \ref{tab:202122}), the highest percentage of cumulative illness-related excess mortality was observed in the range of 52.1\% to 59.7\% for individuals between the ages of 30 and 69. Accordingly, based on the percentage of illness-related excess deaths, it can be concluded that the age groups between 30 to 69 were the most affected. \medskip 

It is noteworthy that for ages 0 to 10, the excess mortality was essentially negative, as illustrated in Figure \ref{fig:excesos202122}b. In the age group of 6 to 10 years, gender-based differences in the pattern were observed. Mortality increased among males but not among females. However, after the age of 10, excess mortality due to illness became positive in both sexes, with a more pronounced impact on males than on females. This observed pattern emphasizes the importance of considering age and gender differentials when assessing the impact of illness-related excess mortality across different age groups. \medskip

The relative excess mortality due to illness displayed a divergent pattern compared to absolute numbers. Notably, the 20 to 29 age group exhibited a percentage of excess mortality similar to that observed in the group of individuals aged 70 and older. In contrast, the 30 to 69 age group showed a comparable percentage of excess mortality between males and females, as illustrated in Figure \ref{fig:excesos202122}b. This discrepancy underscores the importance of considering both absolute numbers and relative percentages when assessing the impact of illness-related excess mortality, revealing nuances in the distribution and impact across different age groups. \medskip

Finally, it's noteworthy that the illness-related excess mortality rate was similar in males and females in the age group under 6 years old and above 11 years old, as evident when comparing Tables \ref{tab:202122hom} and \ref{tab:202122muj}. However, among the age groups of 20 years and older, the illness-related excess mortality rate was consistently between $1.7$ to $2.4$ times higher in males than in females. This gender-based variation highlights age-specific differences in the impact of illness-related excess mortality, emphasizing the need for nuanced analyses across distinct demographic categories. \medskip


\section{Discussion}

Various approaches have been employed to calculate excess deaths during the recent COVID-19 pandemic, as observed in studies such as \cite{Levitt2022, INEGI2023a, Dahal2021, Shkolnikov2021}. These methodologies encompass models based on static averages, linear regression, endemic channels, Serfling regression, splines, Poisson and quasi-Poisson models, negative binomial, generalized linear models, and machine learning. The present work contributes to this array of modeling alternatives by introducing the use of polynomial regression to estimate illness-related excess mortality in Mexico for the COVID-19 epidemic period from 2020 to 2022. This novel approach aims to provide a nuanced perspective on the dynamics of excess mortality and enhance the toolkit of methodologies available for understanding and quantifying the impact of the pandemic on mortality patterns. \medskip 

In Mexico for the year 2020, earlier estimates indicated an excess mortality ranging between $261$ to $276$ per $100,000$ inhabitants. Furthermore, a previously published mortality ratio between males and females was reported as 1.8 times \cite{Dahal2021, AntonioVilla2022}. These estimations closely align with the findings of the present work, which reports an excess mortality rate of $281$ per $100,000$  (refer to Table \ref{tab:2020}). This consistency in results strengthens the robustness and reliability of the excess mortality estimates, as multiple approaches converge on similar figures, contributing to a more comprehensive understanding of the impact of the COVID-19 pandemic on mortality in Mexico during that period. \medskip  

For the two-year period 2020-2021 in Mexico, significant discrepancies exist among various estimates \cite{Wang2022} estimated a total of $798,000$ excess deaths with a 95\% confidence interval (C.I) of $(741,000-867,000)$), while \cite{Palacio2022} estimated $600,590$ with a 95\% C.I $(575,125-626,056)$. The official agency for national statistics in Mexico (INEGI) \cite{INEGI2022} provides two different estimates using distinct models: $636,820$ with a quasi-Poisson model and $704,358$ using endemic channels estimation. Under our proposed methodology, we obtained a point estimation of $719,649$ with a 95\% C.I $(713,240-726,058)$. These variations underscore the methodological diversity and challenges in estimating excess mortality, emphasizing the importance of transparency and understanding the nuances of each approach. \medskip

The data obtained in our study is similar to that of \cite{Wang2022}, but the discrepancy in magnitude can be attributed to two factors. Firstly, we utilized a recently updated database, and secondly, we excluded deaths not caused by diseases. In contrast, our study reports a higher excess than that of \cite{Palacio2022}, but falls within the range of the two estimations reported by INEGI \cite{INEGI2022}. Notably, our methodology yields narrower confidence intervals compared to other studies. This suggests that while our overall estimates align with certain studies, the differences in data sources, inclusion criteria, and modeling techniques contribute to variations in the magnitude of excess mortality estimates across different studies.\medskip

The most recent official report by INEGI \cite{INEGI2023a} estimated for the whole period 2020-2022 an excess mortality by all causes of $673,211$  deaths with a quasi-Poisson model and $807,720$ using endemic channels estimation. In contrast, our estimation of excess deaths is $787,753$, as indicated in Table \ref{tab:202122}. Therefore, our results align more closely with the estimation based on the endemic channel than with the quasi-Poisson model. This highlights the importance of considering and understanding the implications of the chosen modeling approach, as different methodologies may yield varying results in the estimation of excess mortality. \medskip

When comparing the excess mortality rates in 2020 versus 2021 (refer to Tables \ref{tab:2020} and \ref{tab:2021}), the results suggest that the pandemic mitigation strategy implemented in Mexico during the second year had no significant effect on reducing mortality. These findings underscore the disproportionate impact of the pandemic on specific age groups, emphasizing the need for targeted strategies and interventions to mitigate mortality risks in the future. The lack of a substantial reduction in excess mortality rates between the two years may signal the challenges in implementing effective measures or the evolving nature of the pandemic's impact on different demographic cohorts. \medskip

Excess mortality during a pandemic cannot be solely attributed to the direct effect of the new pathogen. It is also necessary to consider the secondary or collateral effects that the pandemic has on healthcare services. For instance, in Mexico, it has been reported that during the first nine months of the pandemic, the country’s primary healthcare system, the Mexican Social Security Institute (IMSS), ceased providing $8.7$ million consultations \cite{Doubova2021}. The utilization of these services began to recover until 2021, particularly concerning conditions like diabetes mellitus and arterial hypertension \cite{Doubova2022}. Other authors have noted that, in 2020, only $34\%$ of individuals in need of medical attention utilized healthcare services \cite{Colchero2021}. Therefore, both hospital re-conversion and the reduction of primary healthcare services were factors that contributed to the increase in excess mortality due to illness during the pandemic in Mexico. These findings underscore the complex interplay between the direct and indirect impacts of a pandemic on healthcare systems and mortality outcomes.\medskip

Our work constitutes an updated and more precise analysis of excess mortality related to the COVID-19 pandemic between 2020 and 2022 in Mexico. This analysis was conducted using a novel statistical approach, enabling a more accurate evaluation of the impact of this pandemic. The refined methodology employed in our study contributes to a nuanced understanding of mortality patterns during this period, providing valuable insights into the complex dynamics of the pandemic's impact on public health. \medskip 

Our results can be summarized in the following key points. First, both absolute and relative excess mortality due to illness were higher in men than in women. Second, mortality rates in women were comparable to those in men, albeit with an age-group difference of approximately ten years. Third, a reduction in excess mortality due to illness was observed in the younger population, particularly among women. Fourth, the excess of mortality during the first two years remained similar. Fifth, the highest relative excess of mortality was detected in the more productive age groups (20 to 69 years). Finally, a substantial reduction in mortality excess was observed until the third year of the pandemic. Therefore, the year of the pandemic, age, and sex were pivotal factors influencing the magnitude of excess mortality due to illness. These findings contribute to a comprehensive understanding of the nuanced dynamics of excess mortality, shedding light on key demographic and temporal factors that played a crucial role during the studied period. \medskip 


\section*{Reproducibility}

All the Julia \cite{Julia2023} programming code, data sets and generated figures and tables are available for reproducibility of results at \url{https://github.com/aerdely/excessmortmx}

\section*{Declaration of interest}

None of the authors have conflicts of interests to declare.

\section*{Ethical statement}

The study uses publicly available data, and as such it did not require ethical approval.


\end{document}